\documentclass[aps,pre,10pt,showpacs,floatfix,onecolumn,nofootinbib,a4paper]{revtex4-2}
\usepackage{amssymb, amsmath, amsthm, mathtools}
\usepackage{bm}
\usepackage{mathrsfs}
\usepackage{hyperref,bookmark}
\usepackage{graphicx}
\usepackage{epsfig,color}
\usepackage{mathrsfs}
\usepackage{verbatim}
\usepackage{url}
\usepackage{subcaption}
\usepackage{float}
\usepackage[greek,english]{babel}
\usepackage{dcolumn}

\newcommand{\tr}{\operatorname{tr}}
\newcommand{\dd}{\operatorname{d}\!}

\newcommand{\diver}{\operatorname{div}}
\newcommand{\curl}{\operatorname{curl}}

\newcommand{\n}{\bm{n}}
\newcommand{\e}{\bm{e}}

\newcommand{\normal}{\bm{\nu}}
\newcommand{\body}{\mathscr{B}}
\newcommand{\free}{\mathscr{F}}
\newcommand{\boundary}{\partial\mathscr{B}}
\newcommand{\Req}{R_\mathrm{e}}

\newcommand{\Q}{\mathbf{Q}}

\newcommand{\arctanh}{\operatorname{arctanh}}

\newcommand{\trans}{^\mathsf{T}}
\newcommand{\Free}{\mathsf{F}}
\newcommand{\Gee}{\mathsf{G}}
\newcommand{\WOF}{W_\mathrm{OF}}
\newcommand{\bend}{\bm{b}}

\newcommand{\Wn}{\mathbf{W}(\n)}
\newcommand{\Pn}{\mathbf{P}(\n)}
\newcommand{\Dn}{\mathbf{D}}
\newcommand{\I}{\mathbf{I}}
\newcommand{\twon}{(\n_1,\n_2)}
\newcommand{\zero}{\bm{0}}

\begin{document}
\latintext

\title{Paradoxes for Chromonic Liquid Crystal Droplets}
\author{Silvia Paparini}
\email{silvia.paparini@unipv.it}
\author{Epifanio G. Virga}
\email{eg.virga@unipv.it}
\affiliation{Dipartimento di Matematica, Universit\`a di Pavia, Via Ferrata 5, 27100 Pavia, Italy}
\begin{abstract}
Chromonic liquid crystals constitute a novel lyotropic phase, whose elastic properties have so far been modeled within the classical Oseen-Frank theory, provided that the twist constant is assumed to be considerably smaller than the saddle-splay constant, in violation of one Ericksen inequality. This paper shows that paradoxical consequences follow from such a violation for droplets of these materials surrounded by an isotropic fluid. For example, tactoids with a degenerate planar anchoring simply disintegrate indefinitely  in myriads of smaller ones.
\end{abstract}
\date{\today}

\maketitle

\section{Introduction}\label{sec:intro}
Chromonic liquid crystals (CLCs) are lyotropic materials, which include Sunset Yellow (SSY), a popular dye in food industry, and disodium cromoglycate (DSCG), an anti-asthmatic drug. In these materials, molecules stuck themselves in columns, which in aqueous solutions develop a nematic orientational order, described by a unit vector field $\n$, called the \emph{director}, representing the average direction in space of the constituting  supra-molecular aggregates. A number of reviews have already appeared in the literature \cite{lydon:chromonic_1998,lydon:handbook,lydon:chromonic_2010,lydon:chromonic,dierking:novel}, to which we refer the reader.

Experiments have been performed with these materials in capillary tubes, with either circular \cite{nayani:spontaneous,davidson:chiral} or rectangular \cite{fu:spontaneous} cross-sections, as well as on cylindrical shells \cite{javadi:cylindrical}, all enforcing \emph{degenerate planar} anchoring, which allows constituting columns to glide freely on the anchoring surface, provided they remain tangent to it. These experiments revealed a tendency of CLCs to acquire in cylinders a \emph{twisted} configuration  at equilibrium, which is represented by an \emph{escaped twist} (ET) director field.

Despite the lack of uniformity in the ground state of these phases \cite{virga:uniform}, they have been modeled by the classical Oseen-Frank theory of nematic liquid crystals, albeit with an anomalously small twist constant $K_{22}$. To accommodate the experimental findings and justify the twisted ground state, this constant has to be smaller than the saddle-splay constant $K_{24}$, in violation of one of the inequalities Ericksen~\cite{ericksen:inequalities} had put forward to guarantee that the Oseen-Frank free energy density be bounded below.

Actually, as shown in \cite{paparini:stability}, such a violation does not prevent the twisted ground state from being locally stable in a cylinder enforcing degenerate planar anchoring. The same reassuring conclusion was reached in \cite{long:violation}. But the question remained as to whether different boundary conditions, still physically significant, could unleash the unboundedness of the total free energy potentially related to the violation of one Ericksen inequality (see also \cite{long:violation} in this connection).

In this paper, we answer this question for the \emph{positive}. If $K_{22}<K_{24}$, a CLC droplet tactoidal\footnote{\emph{Tactoids} are elongated, cylindrically symmetric shapes with pointed ends as poles.} in shape and surrounded by an isotropic fluid environment enforcing degenerate planar anchoring for the director is predicted to be unstable against \emph{shape change}: it would split indefinitely in smaller tactoids while the total free energy plummets to negative infinity.

A similar paradoxical behavior is expected if the splay constant $K_{11}$ is anomalously small. If $K_{11}<K_{24}$, in violation of another Ericksen inequality, a spherical CLC droplet surrounded by a fluid environment enforcing  \emph{homeotropic}\footnote{That is, with $\n$ along the outer unit normal.} anchoring  would split indefinitely in smaller spherical droplets, while the total free energy diverges to negative infinity.

The paper is organized as follows. In Sec.~\ref{sec:theory} we recall a \emph{modicum} of the classical Oseen-Frank theory for nematic liquid crystals, including all Ericksen inequalities. In Sec.~\ref{sec:free_boundary_problem}, we set the scene for the free-boundary problem that need be solved to identify the sequences of shapes that fragment a parent drop proving it unstable. Sections \ref{sec:disintegration_I} and \ref{sec:disintegration_II} are devoted to the explicit construction of such sequences for cases in which one or the other of two Ericksen inequalities are violated. In Sec.~\ref{sec:conclusion}, we draw the conclusions of this work, casting severe doubts on the applicability of the Oseen-Frank theory to describe the elasticity of CLCs. The paper is closed by four Appendices, where a number of mathematical proofs are relegated to ease reading the main text. 

\section{Classical Elastic Theory}\label{sec:theory}
The classical elastic theory of liquid crystals goes back to the pioneering works of Oseen~\cite{oseen:theory} and Frank~\cite{frank:theory}.\footnote{Also a paper by Zocher~\cite{zocher:effect}, mainly concerned with the effect of a magnetic field on director distortions, is often mentioned among the founding contributions. Some go to the extent of also naming the theory after him. Others, in contrast, name the theory only after Frank, as they only deem his contribution to be fully aware of the nature of $\n$ as a \emph{mesoscopic}  descriptor of molecular order.} This theory is variational in nature, as it is based on a bulk free energy functional $\free_\mathrm{b}$ written in the form
\begin{equation}
	\label{eq:free_energy}
	\free_\mathrm{b}[\n]:=\int_{\body}\WOF(\n,\nabla\n)\dd V,
\end{equation}
where $\body$ is a region in space occupied by the material and $V$ is the volume measure. In \eqref{eq:free_energy}, $\WOF$ measures the distortional cost produced by a deviation from a uniform director field $\n$. It is chosen to be the most general frame-indifferent,\footnote{A function $W(\n,\nabla\n)$ is \emph{frame-indifferent} if it is invariant under the action of the orthogonal group $\mathsf{O}(3)$, that is, if $W(\Q\n,\Q(\nabla\n)\Q\trans)=W(\n,\nabla\n)$ for all $\Q\in\mathsf{O}(3)$, where $\Q\trans$ denotes the transpose of $\Q$.}  even function quadratic in $\nabla\n$, 
\begin{equation}
	\label{eq:free_energy_density}
	\WOF(\n,\nabla\n):=\frac{1}{2}K_{11}\left(\diver\n\right)^2+\frac{1}{2}K_{22}\left(\n\cdot\curl\n\right)^2+ \frac{1}{2}K_{33}|\n\times\curl\n|^{2} + K_{24}\left[\tr(\nabla\n)^{2}-(\diver\n)^{2}\right].
\end{equation}
Here $K_{11}$, $K_{22}$, $K_{33}$, and $K_{24}$ are elastic constants characteristic of the material. They are often referred to as the \emph{splay}, \emph{twist}, \emph{bend}, and \emph{saddle-splay} constants, respectively, by the features of the different orientation fields, each with a distortion energy proportional to a single term in \eqref{eq:free_energy_density} (see, for example, Ch.~3 of \cite{virga:variational}). 

Recently, Selinger~\cite{selinger:interpretation} has reinterpreted the classical formula \eqref{eq:free_energy_density} by decomposing the saddle-splay mode into a set of other independent modes. The starting point of this decomposition is a novel representation of $\nabla\n$ (see also \cite{machon:umbilic}),
\begin{equation}
	\label{eq:nabla_n_novel}
	\nabla\n=-\bend\otimes\n+\frac12T\Wn+\frac12S\Pn+\Dn,
\end{equation}
where $\bend:=-(\nabla\n)\n=\n\times\curl\n$ is the \emph{bend} vector, $T:=\n\cdot\curl\n$ is the \emph{twist}, $S:=\diver\n$ is the \emph{splay}, $\Wn$ is the skew-symmetric tensor that has $\n$ as axial vector, $\Pn:=\I-\n\otimes\n$ is the projection onto the plane orthogonal to $\n$, and $\Dn$ is a symmetric tensor such that $\Dn\n=\zero$ and $\tr\Dn=0$. By its own definition, $\Dn\neq\zero$ admits the following biaxial representation,
\begin{equation}
	\label{eq:D_representation}
	\Dn=q(\n_1\otimes\n_1-\n_2\otimes\n_2),
\end{equation}
where $q>0$ and $\twon$ is a pair of orthogonal unit vectors in the plane orthogonal to $\n$, oriented so that $\n=\n_1\times\n_2$.\footnote{It is argued in \cite{selinger:director} that $q$ should be given the name \emph{tetrahedral} splay, to which we would actually prefer \emph{octupolar} splay for the role played by a cubic (octupolar) potential on the unit sphere \cite{pedrini:liquid} in representing all scalar measures of distortion,  but $T$.}

By use of the following identity, 
\begin{equation}
	\label{eq:identity}
	2q^2=\tr(\nabla\n)^2+\frac12T^2-\frac12S^2,
\end{equation}
we can easily give \eqref{eq:free_energy_density} the equivalent form
\begin{equation}
	\label{eq:Frank_equivalent}
	\WOF(\n,\nabla\n)=\frac12(K_{11}-K_{24})S^2+\frac12(K_{22}-K_{24})T^2+\frac12K_{33}B^2+2K_{24}q^2,
\end{equation}
where $B^2:=\bend\cdot\bend$. Since $(S,T,B,q)$ are all independent \emph{distortion characteristics}, it readily follows from \eqref{eq:Frank_equivalent} that $\WOF$ is positive semi-definite whenever
\begin{eqnarray}
	\label{eq:Ericksen_inequalities}
	K_{11}\geqq K_{24}\geqq0,\quad K_{22}\geqq K_{24}\geqq0,\quad K_{33}\geqq0,
\end{eqnarray}
which are the celebrated \emph{Ericksen's inequalities} \cite{ericksen:inequalities}. If these inequalities are satisfied in strict form, the global ground state of $\WOF$ is attained on the uniform director field, characterized by
\begin{equation}
	\label{eq:uniform_ground_state}
	S=T=B=q=0.
\end{equation}

As already mentioned in the Introduction, the third inequality in \eqref{eq:Ericksen_inequalities} must be violated for the ground state of $\WOF$ to be different from \eqref{eq:uniform_ground_state}, involving a non-vanishing $T$. We shall see below how such a choice entails paradoxical consequences. 

Liquid crystals are (within good approximation) incompressible fluids. Thus, when the region $\body$ is \emph{not} fixed, as in the cases considered in this paper, for a given amount of material, $\body$ is subject to
the  \emph{isoperimetric} constraint that prescribes its volume,
\begin{equation}
	\label{eq:isoperimentric_constraint}
	V(\body)=V_0.
\end{equation}
When $\body$ is surrounded by an isotropic fluid, a surface energy arises at the free interface $\boundary$, which, following \cite{rapini:distortion}, we represent as
\begin{equation}
	\label{eq:surface_energy}
	\free_\mathrm{s}[\n]:=\int_{\boundary}\gamma[1+\omega(\n\cdot\normal)^2]\dd A,
\end{equation}
where $\normal$ is the outer unit normal to $\boundary$, $\gamma>0$ is the \emph{isotropic} surface tension, and $\omega>-1$ is a dimensionless parameter bearing the \emph{anisotropic} component of surface tension. For $\omega>0$, $\free_\mathrm{s}$ promotes the \emph{degenerate planar}  anchoring, whereas for $\omega<0$, it promotes the \emph{homeotropic} anchoring. The total free energy functional will then be written as
\begin{eqnarray}\label{eq:free_energy_total}
	\free[\n]:=\free_\mathrm{b}[\n]+\free_\mathrm{s}[\n].
\end{eqnarray}

\section{Free-Boundary Problem}\label{sec:free_boundary_problem}
\begin{figure}[h]
	\centering
	\includegraphics[width=0.33\linewidth]{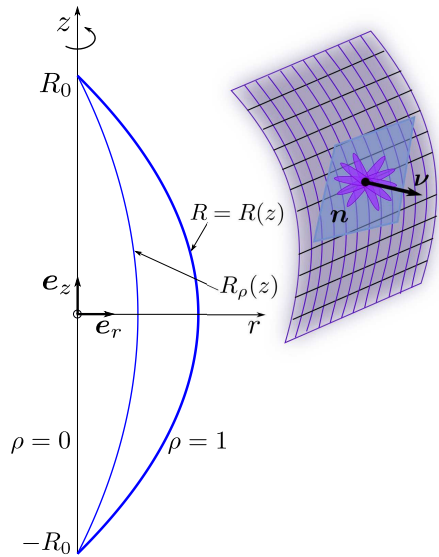}
	\caption{The function $R(z)$ represents the boundary $\boundary$, while $R_\rho(z)=\rho R(z)$, for $\rho\in[0,1]$, is the retraction of $R(z)$, representing a surface on which the polar angle $\beta$ is constant. The director field is tangent to the boundary, but free to orient itself in any direction, as illustrated by the sketch on the  side.} 
	\label{fig:profile}
\end{figure}

A drop comprising a given quantity of CLC is free to adjust its shape $\body$ when surrounded by an isotropic environment, subject only to  \eqref{eq:isoperimentric_constraint}.
In particular, we assume that $\body$ is a region in three-dimensional space rotationally symmetric about the $z$-axis of a standard cylindrical frame $(\e_r, \e_\vartheta, \e_z)$. As shown in Fig. \ref{fig:profile}, the boundary $\boundary$ is obtained by rotating the graph of a smooth function, $R=R(z)$, which  represents the radius of the drop's cross-section at height $z\in\left[-R_0,R_0\right]$. 
The function $R$ vanishes at $z=\pm R_0$, where the drop has its  \emph{poles}. As in \cite{paparini:nematic} (see, for example, equations (19) and (20)), 
the volume of  $\body$ can  be expressed in terms of the function $R(z)$ as 
\begin{equation}
\label{eq:volume_R}
V(\body)=\pi\int_{-R_0}^{R_0}R^2(z)\dd z,
\end{equation}
and the area of  $\boundary$ as
\begin{equation}
\label{eq:area_R}
A(\boundary)=\pi\int_{-R_0}^{R_0}R(z)\sqrt{1+R'(z)^2}\dd z.
\end{equation}
Here and below, a prime $'$ will denote differentiation.

The only requirement for the director $\n$ at the free surface of the drop is to fulfill the \emph{degenerate planar} condition,
\begin{equation}
\label{eq:boundary_planar}
\n|_{\boundary}\cdot\bm{\nu}=0,
\end{equation}
which here, for simplicity, is imposed as a constraint.
As a consequence of \eqref{eq:boundary_planar}, the surface free energy $\free_\mathrm{s}$ reduces to
\begin{equation}
	\label{eq:surface_energy_reduced}
	\free_\mathrm{s}[\n]=\gamma A(\boundary).
\end{equation}
In the present setting, $A(\boundary)$ is given by \eqref{eq:area_R} and $\normal$ is written as
\begin{equation}
	\label{eq:normal_boundary}
	\bm{\nu}=\frac{\e_r-R'\e_z}{\sqrt{1+R'^2}}.
\end{equation}

\begin{figure}[h]
	\centering
	\includegraphics[width=0.15\linewidth]{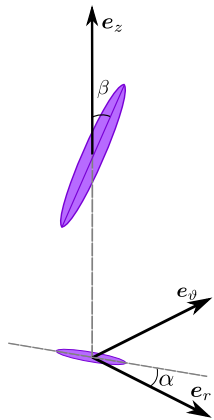}
	\caption{The director field $\n$ is described by the azimuthal angle $\alpha$, which the projection of $\n$ on the $(r,\vartheta)$ plane makes with $\e_r$, and the polar angle $\beta$, which $\n$ makes with the drop's symmetry axis $\e_z$.}
	\label{fig:director_3d}
\end{figure}

For $\n$ to be tangent to $\boundary$, it should be  allowed  to flip out the plane $(\e_\vartheta,\e_z)$. The class of admissible director fields  will thus be described by
\begin{equation}
\label{eq:n_bur_degenerate}
\n=\cos\alpha\sin\beta\e_r+\sin\alpha\sin\beta\e_\vartheta+\cos\beta\e_z,
\end{equation}
where $\alpha\in[0,2\pi)$ is the \emph{azimuthal} angle and $\beta\in[0,\pi]$ is the \emph{polar} angle (see Fig.~\ref{fig:director_3d}).
Here, we shall assume that $\alpha$ depends only on $z$, $\alpha=\alpha(z)$, while $\beta$ depends on both $r$ and $z$, but only through the ratio
\begin{equation}
\label{eq:rho_def_z}
\rho:=\frac{r}{R(z)}\in[0,1].
\end{equation}
The rationale behind this choice is to let $\beta$ be constant on $\boundary$, where $r=R(z)$ and $\rho=1$, and on all surfaces in the interior of $\body$ obtained from $\boundary$ by a \emph{linear retraction} towards the symmetry axis $z$, represented by $R_\rho(z):=\rho R(z)$ with $0\leqq\rho<1$ (see Fig.~\ref{fig:profile}). By letting $\beta=\beta(\rho)$, we assign a polar angle to each retracted surface, the value on one surface being possibly different from the value on other surfaces. Since all retracted surfaces fill the drop, the director field $\n$ is defined on the whole of $\body$ through only two scalar-valued functions in a single variable, $\alpha(z)$ and $\beta(\rho)$.
The constraint in \eqref{eq:boundary_planar} makes these functions not independent, as with the aid of \eqref{eq:normal_boundary} we see from \eqref{eq:n_bur_degenerate} that \eqref{eq:boundary_planar} is valid only if
\begin{equation}
\label{eq:angle_alpha_cos}
\cos\alpha(z)=\frac{R'(z)}{\tan\beta(1)},
\end{equation}
which amounts to the alternative,
\begin{equation}
\label{eq:angle_alpha}
\alpha(z)=\begin{cases}
\arccos\left(\dfrac{R'(z)}{\tan\beta(1)}\right),\\\\
2\pi-\arccos\left(\dfrac{R'(z)}{\tan\beta(1)}\right),
\end{cases}
\end{equation}
whose meaning will soon become clear. Meanwhile, we note that 
a new constraint arises from \eqref{eq:angle_alpha_cos} for $R'$, that is,
\begin{equation}
\label{eq:Rprime_constraint}
 -|\tan\beta(1)|\leqq R'(z)\leqq |\tan\beta(1)|.
\end{equation}

\begin{figure}[h]
	\centering
	\includegraphics[width=0.15\linewidth]{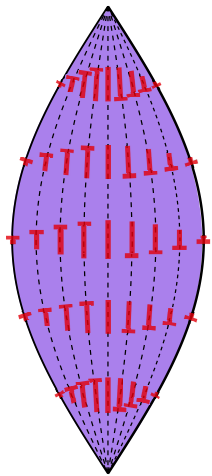}
	\caption{A tactoid with a twisted nematic director field represented as in \eqref{eq:n_bur_degenerate} and \eqref{eq:angle_alpha}. The $(r,z)$ plane of the drawing is a symmetry plane of the drop through its axis. A (red) segment represents $\n$ when it lies on the plane of the drawing, while a nail is used for the projection of $\n$ on that plane when the director is askew with it, the head designating conventionally the end on the same side as the viewer. Dashed lines are cross-sections of the surfaces representing retractions of the boundary.}
	\label{fig:tactoid}
\end{figure}

Figure~\ref{fig:tactoid} illustrates our construction for a \emph{twisted tactoid}: it shows a meridian cross-section of the drop with its family of retracted surfaces to which $\n$ is everywhere tangent. Generically, the director does not lie on the plane of the drawing (spanned by $\e_r$ and $\e_z$), as indicated by the nail symbols, whose heads are conventionally above that plane. 
Neither is the projection of $\n$ on the $(r,z)$ plane tangent everywhere to the lines representing the retracted boundary for $0<\rho<1$. This is confirmed by a careful inspection of Fig.~\ref{fig:tactoid}, but is perhaps better revealed by noting that the outer unit normal $\normal_\rho$ to the retracted surface represented by $R_\rho$ is
\begin{equation}
	\label{eq:retracted_normal}
	\normal_\rho=\frac{\e_r-R_\rho'(z)\e_z}{\sqrt{1+R_\rho'^2}}
\end{equation}
and that, consequently,
\begin{equation}
	\n\cdot\normal_\rho=\frac{R'(z)\cos\beta(\rho)}{\sqrt{1+R_\rho'^2(z)}}\left(\frac{\tan\beta(\rho)}{\tan\beta(1)}-\rho\right).
\end{equation}
In particular, the latter formula shows that the nails depicted in Fig.~\ref{fig:tactoid} are tangent to the cross-sections of the surfaces retracting the boundary on the symmetry plane $z=0$ and, of course, for $\rho=1$ and $\rho=0$ (if, by symmetry, we assume that $\tan\beta(0)=0$). Finally,
choosing one instead of the other alternative in \eqref{eq:angle_alpha} amounts to swap head and tails in the nails. 

Standard computations (deferred to Appendix \ref{sec:computations}) show that the distortion characteristics associated with the field in \eqref{eq:n_bur_degenerate} are given by
\begin{subequations}
\label{distortion_general}
\begin{align}
S=&\frac{1}{R}\left(\cos\alpha\cos\beta\beta'+\frac{1}{\rho}\cos\alpha\sin\beta+\rho R'\sin\beta\beta'\right),\\
T=&\frac{1}{R}\left[\sin\alpha\left(\beta'+\frac{1}{\rho}\cos\beta\sin\beta\right)-\sin^2\beta\alpha' R\right],\label{eq:characteristics_twist}\\
B^2=&\frac{1}{R^2}\left[\beta'^2\left(\rho R'\cos\beta-\cos\alpha\sin\beta\right)^2+\sin^2\beta\left(\alpha'R\cos\beta+\frac{1}{\rho}\sin\alpha\sin\beta\right)\right],\\
2q^2=&\frac{1}{R^2}\left\{\frac{\beta'^2}{2}\left[\left(\cos\alpha\cos\beta+\rho R'\sin\beta\right)^2+\sin^2\alpha\right]+\frac{1}{2\rho^2}\sin^2\beta\left(1-\sin^2\alpha\sin^2\beta\right)\right.\nonumber\\
&\left.-\frac{1}{\rho}\cos\beta\sin\beta\beta'+\frac{1}{2}\left(\alpha' R\right)^2\sin^4\beta+\sin\alpha\sin^2\beta(\alpha'R)\left(\beta'-\cos\beta\sin\beta\right)-R'\cos\alpha\sin^2\beta\beta'\right\}.
\end{align}
\end{subequations}
In particular, \eqref{eq:characteristics_twist} shows that, for given $\beta$, changing the representation of $\alpha$ in accord with \eqref{eq:angle_alpha} only changes the sign of $T$, thus indicating that the two alternative representations of $\alpha$ in \eqref{eq:angle_alpha} correspond to two director fields with opposite \emph{helicity}, that is, winding in opposite senses around the drop's axis. Since both representations for $\alpha$ in \eqref{eq:angle_alpha} are equivalent, hereafter, for definiteness, we shall choose the first, with no prejudice for the generality of our development.

We denote by $\Req$ the \emph{equivalent} radius, that is, the radius  of a  spherical drop with volume $V_0$, and  by $2\mu$ the span (at the poles) of the drop, scaled to the equivalent diameter $2\Req$, 
\begin{equation}
\label{eq:mu_definition_z}
\mu:=\frac{R_0}{\Req}>0.
\end{equation}
We also rescale  lengths $r$, $z$ and $R(z)$ to $\Req$, leaving their names unaltered; with such a renormalization, we further set 
\begin{equation}
	\label{eq:U_xi}
	U(\xi):=\sqrt{\mu}R(z(\xi)),
\end{equation}
where $\xi$ is defined by
\begin{equation}
\label{eq:change_xi}
\xi:=\frac{z}{\mu}\in[-1,1].
\end{equation}
With the aid of \eqref{eq:volume_R}, the volume constraint \eqref{eq:isoperimentric_constraint} then simply reads as 
\begin{equation}
\label{eq:isoperimetric_constraint_xi}
\int_{-1}^{1} U(\xi)^2 \dd \xi=\dfrac{4}{3}.
\end{equation}
For clarity and later use, we record here the form that \eqref{eq:U_xi} acquires in the original dimensional quantities,
\begin{equation}
	\label{eq:U_clarity}
	U\left(\frac{z}{\mu\Req}\right)=\frac{\sqrt{\mu}}{\Req}R(z).
\end{equation}
Furthermore, the function $U$ must vanish at the poles,
\begin{equation}
\label{eq:U_xi_0}
U(\pm1)=0,
\end{equation}
and, by \eqref{eq:Rprime_constraint}, its derivative is subject to the restriction  
\begin{equation}
\label{eq:Rprime_constraint_xi}
-\mu^{3/2}|\tan\beta(1)|\leqq U'(\xi)\leqq<\mu^{3/2}|\tan\beta(1)|.
\end{equation}
For example, for an even, concave function (corresponding to a convex drop $\body$), it would suffice that constraint \eqref{eq:Rprime_constraint_xi} be obeyed for a given   $\mu=\mu_0$ and at $\xi=1$, for it to be valid for all $\mu>\mu_0$ and for all $\xi\in[-1,1]$. On the other hand, smooth shapes are not allowed by \eqref{eq:Rprime_constraint_xi}, as for them $\lim_{\xi\to\pm1}U'(\xi)=\mp\infty$. Thus, hereafter we shall only consider \emph{tactoids}, like the one represented in Fig.~\ref{fig:tactoid}, for which $U'$ is everywhere bounded. 

By use of \eqref{eq:angle_alpha_cos}, \eqref{eq:angle_alpha}, \eqref{distortion_general}, and both changes of variables \eqref{eq:rho_def_z} and \eqref{eq:change_xi}, we give the bulk and surface  free energies the following forms
\begin{subequations}\label{eq:free_energies}
\begin{align}
\label{eq:free_planar}
\mathcal{F}_\mathrm{b}[U,\beta;\mu]&:=\frac{\free_\mathrm{b}[\n]}{2\pi K_{22}\Req}\nonumber\\
&=\frac{1}{\mu^2}\left[\int_{-1}^{1}U'(\xi)^2\dd\xi\right]\mathcal{F}_1[\beta]+\frac{1}{\mu^2}\left[\int_{-1}^{1} \frac{U(\xi)^2U''(\xi)^2}{\mu^3\tan^2\beta(1)-U'(\xi)^2}\dd \xi\right]\mathcal{F}_2[\beta]+\mu\mathcal{F}_3[\beta]
\end{align}
and
\begin{equation}
\label{eq:free_planar_sup}
\mathcal{F}_\mathrm{s}[U;\mu,\upsilon]:=\frac{\free_\mathrm{s}[\n]}{2\pi K_{22}\Req}=\upsilon\sqrt{\mu}\int_{-1}^{1}U(\xi)\sqrt{1+\frac{U'(\xi)^2}{\mu}}\dd\xi,
\end{equation}
\end{subequations}
where
\begin{equation}
	\label{eq:alpha}
	\upsilon:=\frac{\gamma \Req}{K_{22}}
\end{equation}
is a reduced (dimensionless) volume, and the following notation has been 
employed,
\begin{subequations}
\label{eq:element_degenerate}
\begin{align}
\mathcal{F}_1[\beta]&:=\int_0^1\left\{\frac{1}{\tan^2\beta(1)}\left[\frac{\beta'^2}{2}\left(k_1\cos^2\beta+k_3\sin^2\beta-1\right)+\frac{1}{2\rho^2}\left(k_1\sin^2\beta-\cos^2\beta\sin^2\beta+k_3\sin^4\beta\right)\right.\right.\nonumber\\
&\left.\left.\quad\quad\quad+\frac{(k_1-1)}{\rho}\cos\beta\sin\beta\beta'\right] \right.\nonumber\\
&\left.\quad\quad\quad+\frac{1}{\tan\beta(1)}\left[\rho\cos\beta\sin\beta\beta'^2(k_1-k_3)+\sin^2\beta\beta'(k_1-1)+\frac{1}{\rho}\cos\beta\sin^3\beta(k_3-1)\right] \right.\nonumber\\
&\left.\quad\quad\quad+\left[\frac{\rho^2\beta'^2}{2}\left(k_1\sin^2\beta+k_3\cos^2\beta\right)\right]\right\}\rho\dd\rho,\label{eq:F_01}\\
\mathcal{F}_2[\beta]&:=\int_0^1\frac{\sin^2\beta}{2}\left(\sin^2\beta+k_3\cos^2\beta\right)\rho\dd\rho,\label{eq:F_02}\\
\mathcal{F}_3[\beta]&:=\int_0^1\left[\frac{\beta'^2}{2}+\dfrac{1}{2\rho^2}\cos^2\beta\sin^2\beta+\dfrac{k_3}{2\rho^2}\sin^4\beta\right] \rho \dd \rho+\frac{(1-2k_{24})}{2}\sin^2\beta(1),\label{eq:F_mu}
\end{align}
\end{subequations}
which feature the \emph{scaled} elastic constants defined as
\begin{equation}
	\label{eq:elastic_constants_rescaled}
	k_1:=\frac{K_{11}}{K_{22}}, \quad k_3:=\frac{K_{33}}{K_{22}}, \quad k_{24}:=\frac{K_{24}}{K_{22}}.
\end{equation}
The functionals in \eqref{eq:element_degenerate} depend  only on the polar angle $\beta=\beta(\rho)$. For both $\mathcal{F}_1$ and $\mathcal{F}_3$  to be finite, $\beta$ must satisfy the condition
\begin{equation}
	\label{eq:anchoring_director}
	\sin\beta(0)=0,
\end{equation}
which entails that $\n$ is parallel to $\e_z$ along the drop's axis. 
As shown in Appendix \ref{sec:computations}, the dependence of $\mathcal{F}_\mathrm{b}$ on $\alpha$ is hidden in $U'(\xi)$ and $\tan\beta(1)$ through the relation \eqref{eq:angle_alpha_cos} rewritten in the new coordinate $\xi$ in \eqref{eq:change_xi} as
\begin{equation}
\label{eq:angle_alpha_cos_xi}
\cos\alpha(z(\xi))=\frac{U'(\xi)}{\mu^{3/2}\tan\beta(1)}.
\end{equation}

Finally, the appropriate dimensionless form of the total free energy $\free$ in \eqref{eq:free_energy_total} is expressed as the sum of  \eqref{eq:free_planar} and \eqref{eq:free_planar_sup}:
\begin{equation}
\label{eq:free_tot}
\mathcal{F}[U,\beta;\mu,\upsilon]:=\frac{\free[\n]}{2\pi K_{22}\Req}=\mathcal{F}_\mathrm{b}[U,\beta;\mu]+\mathcal{F}_\mathrm{s}[U;\mu,\upsilon],
\end{equation}
where the role of parameters $\mu$ and $\upsilon$ is distinguished from that of functions $U$ and $\beta$ for later convenience.
In the following sections, we shall make use of this expression for $\mathcal{F}$ to show that for either $K_{22}<K_{24}$ or $K_{11}<K_{24}$ there are sequences of droplets, which the drop $\body$ can disintegrate in, so that the total volume $V_0$ is preserved, but the total free energy plummets to $-\infty$. These sequences will provide ground for paradoxes. 


\section{Violating $K_{22}\geqq K_{24}$}\label{sec:disintegration_I}
Here, we construct a family of droplets and associated director fields within a class of distortions with cylindrical symmetry, where the total free energy $\mathcal{F}$ in \eqref{eq:free_tot} does \emph{not} attain a minimum whenever $K_{24}>K_{22}$. We find it convenient to split our discussion into two cases: one where the domain $\body$ can be unbounded and the other where the domain $\body$ is constrained to be bounded.

\subsection{Unconfined Drops}\label{sec:unconfined_drops}
Functional $\mathcal{F}_{3}[\beta]$ in \eqref{eq:F_mu} is nothing but the dimensionless form taken by the Oseen-Frank elastic free energy in a cylinder subject to degenerate planar  boundary conditions \cite{burylov:equilibrium,paparini:stability}. Whenever $k_{24}>1$, the minimizer of $\mathcal{F}_3$ is the escaped twist (ET) field represented by the function 
\begin{equation}
\label{eq:bur_sol}
\beta_{\mathrm{ET}}\left(\rho\right):=\arctan\left(\frac{2\sqrt{k_{24}(k_{24}-1)}\rho}{\sqrt{k_{3}}\left[k_{24}-(k_{24}-1)\rho^2\right]}\right), 
\end{equation}
This field together with its chiral variant, represented by the function  $\widehat{\beta}_{\mathrm{ET}}(\rho):=\pi-\beta_{\mathrm{ET}}(\rho)$, describe the director twist within a CLC tactoid (see Fig.~\ref{fig:tactoid}). They give $\mathcal{F}_3$ one and the same value \cite{paparini:stability},
\begin{equation}
	\label{eq:ET_free_energy}
	\mathcal{F}_3[\beta_{\mathrm{ET}}]=\mathcal{F}_3[\widehat{\beta}_{\mathrm{ET}}]=
	\begin{cases}
		1-k_{24}+\frac{1}{2}\frac{k_3}{\sqrt{1-k_3}}\arctanh\left(\frac{2\sqrt{1-k_3}(k_{24}-1)}{k_3+2(k_{24}-1)}\right), \quad&k_3\leqq1,\\ 1-k_{24}+\frac{1}{2}\frac{k_3}{\sqrt{k_3-1}}\arctan\left(\frac{2\sqrt{k_3-1}(k_{24}-1)}{k_3+2(k_{24}-1)}\right),\quad&k_3\geqq1.
	\end{cases}
\end{equation}
It should be noted that $\mathcal{F}_3[\beta_{\mathrm{ET}}]<0$ and, by \eqref{eq:free_energies}, its contribution to the total free energy $\mathcal{F}$ of the drop in \eqref{eq:free_tot} scales like the dominant power in $\mu$ as $\mu\to\infty$. This suggests a path capable of driving $\mathcal{F}$ to $-\infty$ along a sequence of needle-shaped twisted tactoids, whose polar span grows indefinitely, while the drop's volume is preserved. For this argument to be conclusive, we need to prove that the contributions to $\mathcal{F}$ other than $\mu\mathcal{F}_3$, which are all positive, are unable to counterbalance the divergence of this latter.

To this end, we take $U$ to be a smooth, positive function in the interval $[-1,1]$ , which obeys \eqref{eq:U_xi_0} and \eqref{eq:Rprime_constraint_xi}. Since $U$ is independent of $\mu$, \eqref{eq:Rprime_constraint_xi} is asymptotically satisfied for $\mu\to\infty$, provided that $U'$ is bounded and the integrals in \eqref{eq:free_energies} converge. We can now estimate $\mathcal{F}[U,\beta;\mu]$ for any given $U$ in the above admissible class and $\beta=\beta_{\mathrm{ET}}$ as $\mu\to\infty$. The leading orders in $\mu$ are given by
\begin{equation}
\label{eq:asymptotic_free}
\mathcal{F}[U,\beta_{\mathrm{ET}};\mu,\upsilon]=\mu\mathcal{F}_3[\beta_\mathrm{ET}]+\upsilon\sqrt{\mu}\int_{-1}^1U(\xi)\dd\xi+\mathcal{O}\left(\frac{\upsilon}{\sqrt{\mu}}\right)
\leqq \mu\mathcal{F}_3[\beta_\mathrm{ET}]+\sqrt{\frac{8}{3}}\upsilon\sqrt{\mu}+\mathcal{O}\left(\frac{\upsilon}{\sqrt{\mu}}\right),
\end{equation}
where the inequality follows from H\"older's inequality and \eqref{eq:isoperimetric_constraint_xi}.\footnote{The classical form of H\"older's inequality estimates the integral of $|gf|$, where $f$ and $g$ are functions defined in an real interval (see, for example, \cite[p.\,213]{royden:real}); here it has been applied with $f\equiv1$ and $g=U$.} Thus, for any admissible function $U=U(\xi)$, in the limit as $\mu\to\infty$ the total free energy of a CLC droplet is shown to be unbounded  below whenever $k_{24}>1$. 

Clearly, this disconcerting result revolves about $\mathcal{F}_3[\beta_{\mathrm{ET}}]$ being negative; one could wonder whether adding a constant to the elastic free-energy density $\WOF$ would render $\mathcal{F}_3[\beta_{\mathrm{ET}}]$ positive. It is shown in Appendix \ref{sec:additive_constant} that such a simplistic remedy is indeed illusory. 

\begin{figure}[h]
	\centering
	\includegraphics[width=.5\linewidth]{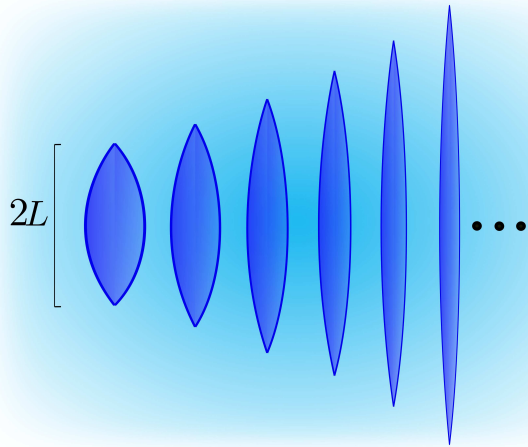}
	\caption{Sequence of tactoidal drops with fixed volume $V_0$ and increasing values of $\mu$, the distance between poles scaled to the diameter of the sphere with equal volume. For $k_{24}>1$, the total free energy diverges to $-\infty$.}
	\label{fig:minimizingsequence}
\end{figure}

Here, we have proved that for $k_{24}>1$ the Oseen-Frank elastic free energy is responsible for the divergence to $-\infty$ of the total free energy of a CLC droplet surrounded by an isotropic fluid enforcing degenerate planar  anchoring on the droplet's free boundary. Our proof is based on the construction of a family of filamentous  twisted tactoids of shape chosen arbitrarily within a wide admissible class (see Fig.~\ref{fig:minimizingsequence}).
For definiteness, in Appendix \ref{sec:R_polynom}, we illustrate the details of this construction for a specific drop's profile.

\subsection{Confined Drops}\label{sec:confined_drops}
An objection could be moved against the disruptive argument presented above: in real life, CLC drops cannot be surrounded by an arbitrarily large amount of fluid, so the minimizing sequence shown in Fig.~\ref{fig:minimizingsequence} would come to a halt as soon as the drop stretches through the largest available length, and no paradox would stand, as the total free energy is finite. Such a pervasive behaviour of CLC drops would be too striking to go unnoticed, but, to the best of our knowledge, it has never been observed. There are, however, also strong theoretical reasons to rebut this objection. They are given here.

\begin{figure}[h] 
	\includegraphics[width=0.66\linewidth]{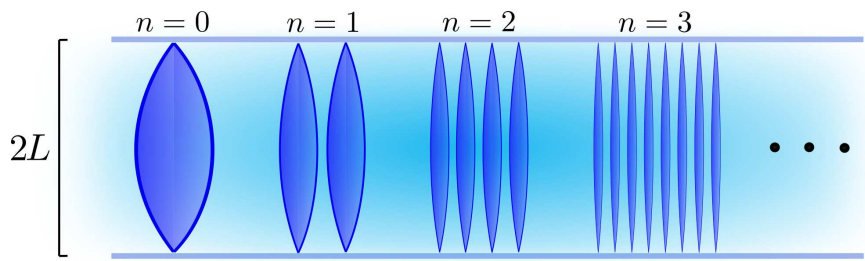}
	\caption{Splitting procedure described in the text: each droplet splits in   halves at every step, thus preserving the total volume.  All drops have one and the same polar span $2L$.}
	\label{fig:crumblingparadox}
\end{figure}

We study the same problem as in Sec.~\ref{sec:unconfined_drops}, but confining drops between two parallel plates, $2L$ apart, and assuming that $2L$ is the maximum polar extension that they have all reached.  We start with a single parent drop of given volume $V_0$ with twisted director field represented by $\beta=\beta_{\mathrm{ET}}$ and boundary profile described by a given function $U(\xi)$ (see Fig.~\ref{fig:crumblingparadox}).
The value of $\mu$ corresponding to this shape is 
\begin{equation}
	\label{eq:mu_L}
	\mu_0:=\frac{L}{\Req},
\end{equation}
for which \eqref{eq:free_tot} delivers a finite (dimensionless)  total free energy $\Free_0$.

We argue that splitting recursively the parent drop in halves, preserving the total volume, will again drive the total free energy to negative infinity. We proceed in steps indexed by the integer $n\in\mathbb{N}$. For $n=1$, the drop is split in two equal parts; for $n=2$, each half is again split in two; etcetera, as shown in Fig.~\ref{fig:crumblingparadox}. The volume of each droplet at step $n$ is $V_n=V_0/2^n$. All droplets have the same polar span $2L$, but since they have different volumes (and so different equivalent radii), the parameter $\mu$ defined by \eqref{eq:mu_definition_z} here depends on $n$ too. The equivalent radius $R_n$ of droplets at step $n$ is  
\begin{equation}
	\label{eq:R_n}
	R_n=2^{-n/3}\Req,
\end{equation}
so that 
\begin{equation}
	\label{eq:mu_n}
	\mu_n:=\frac{L}{R_n}=2^{n/3}\mu_0
\end{equation}
and $\mu_nR_n=\mu_0\Req$ for all $n$. Not only $\mu$, but also $\upsilon$, the reduced volume defined by \eqref{eq:alpha}, depends on $n$:
\begin{equation}
	\label{eq:alpha_n}
	\upsilon_n=\frac{\gamma R_n}{K_{22}}=2^{-n/3}\upsilon_0,
\end{equation}
where $\upsilon_0$ is the reduced volume of the parent drop. 

It should be noted that the splitting strategy adopted here affects the droplet's shape, while leaving the function $U$ unchanged. Keeping in mind that the equivalent radius at step $n$ is $R_n$, we rewrite  $U$ as expressed by \eqref{eq:U_clarity} in  terms of  dimensional $R(z)$,
\begin{equation}
	\label{eq:U_clarity_rewritten}
	U\left(\frac{z}{\mu_n R_n}\right)=\frac{\sqrt{\mu_n}}{R_n}R(z),
\end{equation}
from which, with the aid of \eqref{eq:mu_L}, \eqref{eq:R_n}, and \eqref{eq:mu_n}, it follows that
\begin{equation}
	\label{eq:R_clarity}
	R(z)=2^{-n/2}\Req\sqrt{\frac{\Req}{L}}U\left(\frac{z}{L}\right)\quad\text{for}\quad-L\leqq z\leqq L.
\end{equation}

Similarly, for the total free energy $\Free_n$ at step $n$ (scaled to $2\pi K_{22}\Req$), we readily obtain the estimate
\begin{equation}
	\label{eq:F_n_estimate}
	\Free_n=\frac{R_n}{\Req}2^n\mathcal{F}[U,\beta_{\mathrm{ET}};\mu_n,\upsilon_n]\leqq2^n\mu_0\mathcal{F}_3[\beta_{\mathrm{ET}}]+\sqrt{\frac{8}{3}}2^{n/2}\upsilon_0\sqrt{\mu_0}+\mathcal{O}(2^{-n/2}),\quad n\to\infty.
\end{equation}  
Since $\mathcal{F}_3[\beta_{\mathrm{ET}}]<0$, \eqref{eq:F_n_estimate} implies the divergence of $\Free_n$ to negative infinity as the splitting proceeds indefinitely.

This confirms that the total free energy of a CLC drop is unbounded below also in the confined case. Being, however, an asymptotic argument, it still leaves room for an objection, more of a physical than mathematical nature. For the above splitting strategy to be interpreted as a shape instability for the parent drop, we should prove that
\begin{equation}
	\label{eq:first_step_inequality}
	\Free_1<\Free_0.
\end{equation}
It does not suffice to know that $\Free_n\to-\infty$. Indeed, it is not difficult to show that for \eqref{eq:first_step_inequality} to be valid $\mu_0$ must be sufficiently large. However, as proved in Appendix~\ref{sec:lambda_crit}, if we split the parent drop in appropriate unequal components, we can always guarantee the validity of $\eqref{eq:first_step_inequality}$, thus proving that the violation of Ericksen's inequality $K_{22}\geqq K_{24}$ makes a CLC drop unstable against domain splitting.

In the following section, we shall use a similar argument to show that a spherical drop subject to homeotropic anchoring on its boundary would disintegrate if Ericksen's inequality $K_{11}\geqq K_{24}$ is violated.

\section{Violating $K_{11}\geqq K_{24}$}\label{sec:disintegration_II}
Here we assume that $K_{11}<K_{24}$ and consider a spherical droplet $\body_0$ of volume $V_0$ enforcing \emph{homeotropic anchoring} for the director $\n$ on its boundary,
\begin{equation}\label{eq:homeotropic_anchoring}
\n|_{\partial\body_0}\cdot\normal=1,
\end{equation}
where $\normal$ is the outer unit normal to $\partial\body_0$. The \emph{radial hedgehog} $\n_\mathrm{H}=\e_r$, that is, the director field everywhere directed like the  unit vector field $\e_r$ emanating from the center of $\body_0$, is a \emph{universal} solution \cite{ericksen:general}. It satisfies the equilibrium equations associated with any frame-indifferent  free energy density $W(\n,\nabla\n)$.\footnote{Since the Oseen-Frank energy density $W_\mathrm{OF}$ in \eqref{eq:free_energy_density} is frame-indifferent, $\n_\mathrm{H}$ is an equilibrium solution for $\free$, for any choice of the elastic constants.} Moreover, $\n_\mathrm{H}$ clearly obeys \eqref{eq:homeotropic_anchoring}. With the aid of \eqref{eq:free_energy} and \eqref{eq:surface_energy}, the total free energy functional $\free$ in \eqref{eq:free_energy_total} is readily computed for $\n_\mathrm{H}$; its scaled value is
\begin{equation}
	\label{eq:free_energy_H}
	\mathcal{F}^\ast:=\frac{\free[\n_\mathrm{H}]}{4\pi K_{11}\Req}=1-k_{24}^\ast+\upsilon^\ast(1+\omega),
\end{equation}
where, in analogy with \eqref{eq:elastic_constants_rescaled} and \eqref{eq:alpha}, we have set 
\begin{equation}
	\label{eq:k_24_star}
	k^\ast_{24}:=\frac{K_{24}}{K_{11}}>1\quad\text{and}\quad \upsilon^\ast:=\frac{\gamma\Req}{K_{11}}.
\end{equation}

We now proceed as in Sec.~\ref{sec:confined_drops}, splitting the drop into $2^n$ equal spherical components, each of radius $R_n$ as in \eqref{eq:R_n}. The formula for the total free energy $\Free_n^\ast$ (scaled to $4\pi K_{11}\Req$) that mimics \eqref{eq:F_n_estimate} is here
\begin{equation}\label{eq:free_n_asympt_h}
	\Free^\ast_n=2^{2n/3}[(1-k_{24}^\ast)+2^{-n/3}\upsilon^\ast(1+\omega)]=2^{2n/3}(1-k_{24}^\ast)+\mathcal{O}\left(2^{n/3}\right)\to-\infty,\quad n\to\infty,
\end{equation}
which proves the asymptotic instability of the parent spherical drop when the Ericksen inequality $K_{11}\geqq K_{24}$ is violated.

As already remarked in Sec.~\ref{sec:confined_drops}, this reasoning does not guarantee that the parent spherical drop splits spontaneously in halves. For this to be the case, it must be  
$\Free_1^\ast<\Free_0^\ast$, which  requires that
\begin{equation}
\label{eq:v_max_2_h}
\upsilon^\ast<\frac{(k_{24}^\ast-1)}{(1+\omega)}\frac{(2^{2/3}-1)}{(2^{1/3}-1)},
\end{equation}
thus setting an effective upper bound on the drop's initial volume.

A drop is however unstable also when \eqref{eq:v_max_2_h} is not satisfied. To see this, for given $0<\lambda<\frac12$, we split the parent drop into two \emph{unequal} spherical components, with volumes $V_1$ and $V_2$ adding up to $V_0$,
\begin{equation}
	\label{eq:V1_V2}
	V_1=\lambda V_0, \quad V_2=(1-\lambda)V_0
\end{equation}
and corresponding  radii
\begin{equation}
	\label{eq:R1_R2}
	R_1=\lambda^{1/3} \Req, \quad R_2=(1-\lambda)^{1/3}\Req.
\end{equation}
The total free energy $\Free_\lambda^\ast$ (again scaled  to $4\pi K_{11}\Req$) is now given by
\begin{equation}
	\label{eq:free_lambda_h}
	\Free_\lambda^\ast=\lambda^{1/3}\left[(1-k_{24}^\ast)+\lambda^{1/3}\upsilon^\ast(1+\omega)\right]+(1-\lambda)^{1/3}\left[(1-k_{24}^\ast)+(1-\lambda)^{1/3}\upsilon^\ast(1+\omega)\right]
\end{equation}
and the inequality $\Free_\lambda^\ast<\Free_0^\ast$ is satisfied for
\begin{equation}
	\label{eq:v_max_lambda_h}
	\upsilon^\ast<\frac{(k_{24}^\ast-1)}{(1+\omega)}\frac{\left[\lambda^{1/3}+(1-\lambda)^{1/3}-1\right]}{\left[\lambda^{2/3}+(1-\lambda)^{2/3}-1\right]}= \mathcal{O}\left(\lambda^{-1/3}\right)\to+\infty, \quad \lambda\to0.
\end{equation}
Thus, for every given volume of the drop there is a splitting fraction $\lambda>0$ corresponding to a net decrease in the total free energy.

\section{Conclusions}\label{sec:conclusion}
This paper shows the paradoxical consequences stemming from adopting the classical Oseen-Frank elastic theory to descibe CLCs when either of the following Ericksen inequalities is violated,
\begin{equation}
	\label{eq:Ericksen_inequalities_rewritten}
	K_{22}\geqq K_{24},\quad K_{11}\geqq K_{24}.
\end{equation}
Violation of the former is at the heart of the commonly accepted understanding of CLCs, as it substantiates the experimentally observed ground state of these materials, which in capillary cylinders with degenerate planar boundary conditions take one of two symmetric twisted director configurations, swaying away from the uniform orientation along the cylinder's axis, which is the alignment that ordinary nematics would prefer.

As shown in \cite{paparini:stability}, violation of \eqref{eq:Ericksen_inequalities_rewritten}$_1$ in the presence of degenerate planar anchoring is \emph{not} prejudicial to the stability of CLC's twisted ground state (see also \cite{long:violation}); this has perhaps nurtured the hope that \eqref{eq:Ericksen_inequalities_rewritten}$_1$  may  be renounced in the Oseen-Frank theory of CLCs. Our paper proves that this is not the case, as such a relaxed theory would entail shape instability of tactoids, an instability which, to our knowledge, has not been observed, and which we deem paradoxical.
When \eqref{eq:Ericksen_inequalities_rewritten}$_2$ is violated, a similar shape instability is predicted, this time for spherical droplets with homeotropic anchoring.

We know about a single material for which both inequalities in \eqref{eq:Ericksen_inequalities_rewritten} are allegedly violated. This is 
SSY, for which the following values of the elastic constants were measured in \cite{zhou:elasticity_2012},  $K_{11} = 4.3\mathrm{pN}$, $K_{22} = 0.7\mathrm{pN}$, and $K_{33} = 6.1\mathrm{pN}$ and it was found in \cite{davidson:chiral} that $K_{24}=15.8\mathrm{pN}$. These experimental values were obtained by assuming valid the Oseen-Frank theory, which is precisely the assumption that we contend here. Thus, on logical grounds, one cannot say that our findings are in contrast with the experimental evidence. 

The direct consequence of our study is that a novel elastic theory is in order for CLCs, capable of overcoming the paradoxical conclusions that the Oseen-Frank theory would lead us to. Some timid proposals have already been advanced. For example, in \cite{long:violation} the role of added disclinations is advocated (provided that their energy cost can be made sufficiently low), whereas in \cite{paparini:thesis} a quartic twist term is added to the Oseen-Frank free energy density, which has the potential to restore shape stability when \eqref{eq:Ericksen_inequalities_rewritten}$_1$ is violated  (but \eqref{eq:Ericksen_inequalities_rewritten}$_2$ is not). We are presently pursuing further this line of thought, although we are aware that it may suffer from the many difficulties encountered by other higher-order theories.

\appendix

\section{Useful Computations}\label{sec:computations}
This technical Appendix contains ancillary results used in Section \ref{sec:free_boundary_problem}. For the particular class of distortions described by \eqref{eq:n_bur_degenerate} with $\alpha$ as a smooth function of $z$, we compute
\begin{align}
\label{eq:nabla_n}
\nabla\n=\frac{1}{R}&\Big\{\cos\alpha\cos\beta\beta'\e_r\otimes\e_r-\frac{1}{\rho}\sin\alpha\sin\beta\e_r\otimes\e_\theta-\left(\sin\alpha\sin\beta\alpha'R+\rho R'\cos\alpha\cos\beta\beta'\right)\e_r\otimes\e_z\nonumber\\
&+\sin\alpha\cos\beta\beta'\e_\theta\otimes\e_r+\frac{1}{\rho}\cos\alpha\sin\beta\e_\theta\otimes\e_\theta+\left(\cos\alpha\sin\beta\alpha'R-\rho R'\sin\alpha\cos\beta\beta'\right)\e_\theta\otimes\e_z\nonumber\\
&-\sin\beta\beta'\e_z\otimes\e_r+\rho R'\sin\beta\beta'\e_z\otimes\e_z\Big\},
\end{align}
where, as in the main text, a prime denotes differentiation.

The following identities justify the expression for the reduced functionals \eqref{eq:free_planar} and \eqref{eq:free_planar_sup}; they are obtained making use of  \eqref{eq:angle_alpha_cos},
\begin{subequations}
\label{eq:integrals_alpha}
\begin{align}
\int_{-\mu}^{\mu}\cos^2\alpha\dd z=&\frac{1}{\tan^2\beta(1)}\int_{-\mu}^{\mu}R'^2\dd z=\frac{1}{\mu^2\tan^2\beta(1)}\int_{-1}^{1}U'^2\dd \xi,\\
\int_{-\mu}^{\mu}\sin\alpha\alpha'R\dd z=&-\frac{1}{\tan\beta(1)}\int_{-\mu}^{\mu}R''R\dd z=\frac{1}{\tan\beta(1)}\int_{-\mu}^{\mu}R'^2\dd z=\frac{1}{\mu^2\tan\beta(1)}\int_{-1}^{1}U'^2\dd \xi,\label{eq:int_sinalphaalphaprimeR}\\
\int_{-\mu}^{\mu}\alpha'^2R^2\dd z=&\int_{-\mu}^{\mu}\frac{R''^2R^2}{\tan\beta(1)^2-R'^2}\dd z=\frac{1}{\mu^2}\int_{-1}^{1}\frac{U''^2U^2}{\mu^3\tan\beta(1)^2-U'^2}\dd \xi.
\end{align}
\end{subequations}
Here $\xi$ is the variable defined in \eqref{eq:change_xi} and an integration by parts has been performed in \eqref{eq:int_sinalphaalphaprimeR} with the aid of \eqref{eq:U_xi_0}.

\section{Useless  Constant}\label{sec:additive_constant}
When $k_{24}>1$, the ET configuration \eqref{eq:bur_sol} realizes the minimum of $\mathcal{F}_3[\beta]$, the dimensionless form of the Oseen-Frank elastic free energy in a cylinder subject to degenerate boundary conditions, and possess less elastic free energy than the uniform alignment $\n=\e_z$, described by $\beta\equiv0$, for which $\mathcal{F}_3$ vanishes.

We have seen that the divergence to negative infinity of the functional in \eqref{eq:free_tot} in the sequences of droplets considered in Sec.~\ref{sec:disintegration_II} stems from being $\mathcal{F}_3[\beta_{\mathrm{ET}}]<0$. One could wonder whether the Oseen-Frank energy density $\WOF$ might be altered by an additive constant $c$ chosen so as to make positive the minimum energy of the ET configuration in a cylinder. This question is easily answered for the positive, but it turns  out that $c$ depends on the cylinder's radius $R$,
\begin{equation}
\label{eq:c_add}
c=-\frac{2 K_{22}}{R^2}\mathcal{F}_3[\beta_\mathrm{ET}],
\end{equation}
and, failing to be intrinsic, it is of no use.

\section{Sinusoidal Profile}\label{sec:R_polynom}
\begin{figure}[h] 
	\includegraphics[width=.37\linewidth]{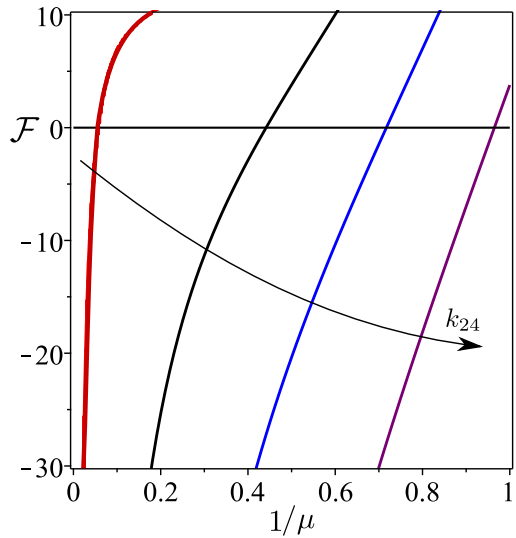}
	\caption{Graphs of $\mathcal{F}$ in \eqref{eq:free_tot} against $1/\mu$ for $U$ as in  \eqref{eq:U_polynomial} and $\beta=\beta_{\mathrm{ET}}$ as in \eqref{eq:bur_sol} for $k_1=k_3=10$, $\upsilon=10$ and a sequence of values of $k_{24}>1$, precisely, $k_{24}=1.2, \, 4., \, 7.5, \, 12$ (arranged in  increasing order, as indicated  by the arrow). Whenever $k_{24}>1$, $\mathcal{F}$ is  unbounded below and diverges to $-\infty$ as $\mu$ tends to $\infty$.}
	\label{fig:plot_F_strong_anchoring}
\end{figure}

We present here  an illustrative example, in which the drop's profile is described by  the following sinusoidal function
\begin{equation}
\label{eq:U_polynomial}
U=\frac{2}{\sqrt{3}}\cos\left(\frac{\pi \xi}{2}\right),
\end{equation}
which vanishes at the poles, where $\xi=\pm1$, and satisfies \eqref{eq:isoperimetric_constraint_xi}. For $U$ as in \eqref{eq:U_polynomial}, 
\eqref{eq:Rprime_constraint_xi} is satisfied whenever
\begin{equation}\label{eq:mu_inequality}
\mu\geqq\frac{1}{3^{1/3}2^{1/3}}\left(\frac{\sqrt{k_3}}{\sqrt{k_{24}(k_{24}-1)}}\right)^{2/3}.
\end{equation}
The functional $\mathcal{F}$ in \eqref{eq:free_tot} has been computed numerically for $U$ as in \eqref{eq:U_polynomial}, $\beta=\beta_{\mathrm{ET}}$, and $\mu$ satisfying \eqref{eq:mu_inequality}. The outcome is illustrated by the graphs shown in
Fig.~\ref{fig:plot_F_strong_anchoring} for $k_1=k_3=10$, $\upsilon=10$, and different values of $k_{24}>1$.
For every $k_{24}>1$, in the limit as $\mu$ tends to $\infty$, $\mathcal{F}$
 does not attain a minimum and diverges to $-\infty$, as expected.

\section{Destabilizing Splitting}\label{sec:lambda_crit}
In this Appendix, we reason as in Sec.~\ref{sec:disintegration_II} to show that when $k_{24}>1$ a confined droplet can be split into two unequal components in such a way that the total free energy is decreased.

Let $0<\lambda<\frac12$ be given and let the volumes of the split droplets $V_1$ and $V_2$ be defined as in \eqref{eq:V1_V2}, for which $R_1$ and $R_2$ in \eqref{eq:R1_R2} play the role of equivalent radii. Correspondingly, there are two parameters defined for each droplet as in \eqref{eq:mu_L},
\begin{equation}
\label{eq:mu1_mu2}
\mu_1=\lambda^{-1/3} \mu_0, \quad \mu_2=(1-\lambda)^{-1/3}\mu_0,
\end{equation}
and two dimensionless volumes, 
\begin{equation}
\label{eq:alpha1_alpha2}
\upsilon_1=\lambda^{1/3} \upsilon, \quad \upsilon_2=(1-\lambda)^{1/3}\upsilon.
\end{equation}

In analogy with \eqref{eq:F_n_estimate}, the total free energy $\Free_\lambda$ of the pair of split droplets (scaled to $2\pi K_{22}\Req$) is given by
\begin{equation}
	\label{eq:F_lambda}
	\Free_\lambda=\lambda^{1/3}\mathcal{F}[U,\beta_{\mathrm{ET}};\mu_1,\upsilon_1]+(1-\lambda^{1/3})\mathcal{F}[U,\beta_{\mathrm{ET}};\mu_2,\upsilon_2],
\end{equation}
where $\mathcal{F}$ is delivered by \eqref{eq:free_tot}. To ease our proof, with the aid of \eqref{eq:free_energies} and \eqref{eq:element_degenerate}, we now rewrite $\mathcal{F}$ as
\begin{equation}
	\label{eq:F_rewritten}
	\mathcal{F}[U,\beta;\mu,\upsilon]=\frac{1}{\mu^2}\Gee_1[U]\mathcal{F}_1[\beta]+\frac{1}{\mu^2}\Gee_2[U;\mu]\mathcal{F}_2[\beta]+\mu\mathcal{F}_3[\beta]+\upsilon\sqrt{\mu}\Gee_\mathrm{s}[U;\mu],
\end{equation}
where we have set
\begin{equation}
\label{eq:functionals_G}
\Gee_1[U]:=\int_{-1}^1U'(\xi)^2\dd\xi,\quad 
\Gee_2[U;\mu]:=\int_{-1}^{1} \frac{U(\xi)^2U''(\xi)^2}{\mu^3\tan^2\beta(1)-U'(\xi)^2}\dd \xi, \quad \Gee_\mathrm{s}[U;\mu]:=\int_{-1}^{1}U(\xi)\sqrt{1+\frac{U'(\xi)^2}{\mu}}.
\end{equation}
Since both $\Gee_2$ and $\Gee_\mathrm{s}$ are monotonically decreasing in $\mu$ and, by \eqref{eq:mu1_mu2}, $\mu_i>\mu_0$, for $i=1,2$, we readily see from \eqref{eq:F_rewritten} that
\begin{align}
	\label{eq:F_lambda_comparison}
	\Free_\lambda&<\frac{1}{\mu_0^2}\Gee_1[U]\mathcal{F}_1[\beta_{\mathrm{ET}}]+\frac{1}{\mu_0^2}\Gee_2[U;\mu_0]\mathcal{F}[\beta_{\mathrm{ET}}]+2\mu_0\mathcal{F}_3[\beta_{\mathrm{ET}}]
	+\upsilon\sqrt{\mu_0}\left[\lambda^{1/2}+(1-\lambda)^{1/2}\right]\Gee_\mathrm{s}[U;\mu_0]\nonumber\\
	&=\Free_0+\mu_0\mathcal{F}_3[\beta_{\mathrm{ET}}]+\upsilon\sqrt{\mu_0}\left[\lambda^{1/2}+(1-\lambda)^{1/2}-1\right]\Gee_\mathrm{s}[U;\mu_0],
\end{align}
where $\Free_0$ is the total free energy of the parent drop. The inequality $\Free_\lambda<\Free_0$ is then valid for
\begin{equation}
\label{eq:upper_bound_v_lambda}
\upsilon<-\frac{\sqrt{\mu_0}}{\left[\lambda^{1/2}+(1-\lambda)^{1/2}-1\right]}\frac{\mathcal{F}_3[\beta_\mathrm{ET}]}{\Gee_\mathrm{s}[U;\mu_0]}=\mathcal{O}\left(\lambda^{-1/2}\right)\to+\infty,\quad\lambda\to0.
\end{equation}
The divergence of this upper bound for $\upsilon$ as   $\lambda$ tends to $0$ guarantees that there is always a $\bar{\lambda}\in(0,\frac12]$ such that for every $\lambda\in(0,\bar\lambda)$ the inequality \eqref{eq:upper_bound_v_lambda} is satisfied for a given $\upsilon$, and so the parent drop is unstable.

%

\end{document}